\documentclass[a4,11pt]{iopart}

\usepackage{amstext}
\usepackage{amssymb}
\usepackage{pstricks}
\usepackage{iopams}
\usepackage{graphics}
\usepackage{graphicx}

\usepackage{epsfig}

\newcommand{\ee}{\text{e}}
\newcommand{\eps}{\varepsilon}
\newcommand{\LDF}{large deviation function }
\newcommand{\C}{{\mathcal C}}
\newcommand{\Y}{{\mathcal Y}}
\newcommand{\NN}{{\mathcal N}}

\begin{document}

\title[A numerical approach to large deviations in continuous-time]
{A numerical approach to large deviations in continuous-time}

\author{Vivien Lecomte$^{1,2}$, Julien Tailleur$^{3}$}

\address{\ $^1$ Laboratoire de Physique Th\'eorique (CNRS UMR 8627), \\
	Universit\'e de Paris XI, 91405 Orsay cedex, France}

\address{\ $^2$ Laboratoire Mati\`ere et Syst\`emes Complexes 
	(CNRS UMR 7057), \\ Universit\'e Paris VII -- Denis Diderot, 
        10 rue Alice Domon et L\'eonie Duquet, \\ 75205 Paris cedex 13, France}

\address{\ $^3$ Laboratoire PMMH (UMR 7636 CNRS, ESPCI, P6, P7) \\
         10 rue  Vauquelin, 75231 Paris cedex 05, France} 

\eads{\mailto{vivien.lecomte@th.u-psud.fr}, \mailto{tailleur@pmmh.espci.fr}}

\begin{abstract}

  We present an algorithm to evaluate large deviation functions
  associated to history-dependent observables. Instead of relying on a
  time discretisation procedure to approximate the dynamics, we
  provide a direct continuous-time algorithm valuable for systems
  with multiple time scales, thus extending the work of Giardin\`a,
  Kurchan and Peliti~\cite{giardinakurchanpeliti}.
  
  The procedure is supplemented with a thermodynamic-integration
  scheme which improves its efficiency.  We also show how the method
  can be used to probe large deviation functions in systems with a
  dynamical phase transition -- revealed in our context through the
  appearance of a non-analyticity in the large deviation functions.

\end{abstract}




\maketitle

\section{Introduction}


The statistical physics of equilibrium systems was first designed to
reproduce the macroscopic predictions of thermodynamics, but it was
soon realised~\cite{einstein05} that it also provides a well-suited
frame to describe the fluctuations of physical observables. Such a
theory is not available when dealing with non-equilibrium systems or
dynamical observables, as one lacks thermodynamics functions such
as the free-energy.
Over the last decade, there has been a growing interest within the 
physics community in the theory of large deviation functions~\cite{ellis85},
as it appeared that they could fill this gap in some cases.
For instance, the fluctuations of particle or energy currents $Q$
flowing through a system in the \emph{long time} limit can be obtained
from the associated large deviation function $\pi(q=Q/t)$:
\begin{equation} \label{eq:pi_of_q}
  \text{Prob}(q \,,\: t) \ \sim\ 
  \ee^{t \pi(q)}
  \qquad \text{as} \quad t\to\infty
\end{equation}
The function $\pi(q)$ is a dynamical analog of the intensive entropy
in the microcanonical ensemble and the long time limit plays the role
of the thermodynamic limit.  In the past few years, a huge amount of
effort has been devoted to the study of a strikingly simple symmetry
of the \LDF of the injected power, the so-called fluctuation theorem.
The results obtained range from theoretical
and numerical studies
\cite{evanscohenmorriss,gallavotticohen,jarzynski,kurchan,crooks,lebowitzspohn}
to experimental applications (see~\cite{ldf-exp} for a glimpse at the
literature). 

Beyond its symmetries, $\pi(q)$ itself expectedly bears information on
the current flowing through the system. For instance, the \LDF
$\pi(q)$ can be fully determined in diffusive
systems~\cite{bodineauderrida,bertinidesole} and its scaling properties differ
from those of superdiffusive ones (see \cite{derridalebowitzappert}
for explicit results).  In the context of dynamical systems also, large deviation functions
associated with more general observables have been
introduced~\cite{ruelle} and their determinations has received a broad
interest~\cite{grassberger,gaspard,thermo-formalism,lyap-weighted-dyn}.
Although the variety of results obtained from these approaches raises hopes
for an out-of-equilibrium thermodynamics, the determination of large
deviation functions is in general a hard task to achieve and most
exact results are confined to simple systems or peculiar
models~\cite{bodineauderrida,bertinidesole,derridalebowitzappert}. In
more complex cases, we have to rely on numerical evaluations.

However, the very definition of large deviations renders their direct
numerical observation almost impossible, as the probability to observe
a value of $Q/t$ far from its average typically decreases
exponentially with time. Giardin\`a, Kurchan and
Peliti~\cite{giardinakurchanpeliti} introduced a numerical procedure
which overcomes this difficulty for discrete time Markov chains.
Nevertheless, most physical processes evolve continuously in time, and
one thus has to choose an arbitrary time step $dt$ to discretise the
dynamics, balancing between algorithm efficiency and errors arising
from the approximation.  Typically, $dt$ has to be smaller than any
time scale of the system, yet too small a value only increases
the simulation duration, since most of the time steps would then be
spent in rejected moves.
This issue already affects the standard Monte Carlo algorithms, and
becomes quite unavoidable when dealing with the computation of large
deviation functions.  Indeed, even systems featuring a single time
scale in the steady state present in general different time scales in
their large deviations, depending on the kind of histories probed,
which makes the choice of $dt$ strenuous.
A simple example is given by traffic flow models, in which the typical
time scale is fixed on average, but vary by a factor
equal to the number $N$ of ``cars'' when comparing jammed histories
(where $\mathcal{O} (1)$ cars move) and free flowing histories (where
$\mathcal{O}(N)$ cars move).

\medskip In the present paper we propose a new procedure where the
time discretisation issue is bypassed with a direct continuous-time
approach.
The outline of the paper is as follows: in section~\ref{sec:algo}, we
recall the continuous-time formalism, define the large deviation
functions and present the algorithm. In section~\ref{sec:examples}, we
study three systems where the continuous-time approach proves useful:
the symmetric simple exclusion process, its asymmetric, out of
equilibrium counterpart, and the contact process, for which a
dynamical phase transition occurs.

\section{Formalism and Algorithm}
\label{sec:algo}

\subsection{Continuous time Markov chains}

\begin{figure}
\begin{center}

\begin{pspicture}(0,-1)(12,1.2)
   \psline{|-}(0, 0)(3,0)
   \psline{|-}(3,0)(5,0)
   \psline{|-}(5,0)(8,0)
   \psline{|-}(8,0)(9,0)
   \psline{|-|}(9,0)(12,0)

   \rput[b](0,0.3){$t_0=0$}
   \rput[t](0,-.3){$\C_0$}
   \rput[b](3,0.3){$t_1$}
   \rput[t](3,-.3){$\C_1$}
   \rput[b](5,0.3){$t_2$}
   \rput[t](5,-.3){$\C_2$}
   \rput[b](6.5,0.3){$\ldots$}
   \rput[b](8,0.3){$t_{K-1}$}
   \rput[t](8,-.3){$\C_{K-1}$}
   \rput[b](9,0.3){$t_K$}
   \rput[t](9,-.3){$\C_K$}
   \rput[b](12,0.3){$t$}
   \rput[t](12,-.3){$\C=\C_K$}

\end{pspicture}

\caption{An history of the system between time $t_{0}=0$ and time $t$.}
\label{fig:timearrow}
\end{center}
\end{figure}
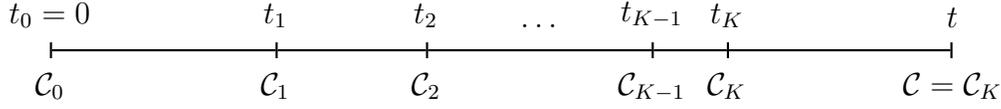

We consider a system described by a finite number of configurations
$\{\C\}$, whose evolution is determined by the transition rates
$W(\C\to\C')$ between different configurations.
The probability $P(\C,t)$ to find the system in $\C$ at time $t$
evolves according to the master equation
\begin{equation} \label{eqn:mait}
  \partial_t P(\C,t) = \sum_{\C'\neq\C} 
  W(\C'\to\C) P(\C',t)\ - r(\C) P(\C,t)
\end{equation}
where the escape rate $r(\C)$ reads
\begin{equation}
  r(\C) = \sum_{\C'\neq\C} W(\C\to\C') .
\end{equation}
Starting from $\C_{0}$, the system evolves through a succession of
configurations $\{\C_k\}_{0\leq k\leq K}$, jumping from $\C_k$ to
$\C_{k+1}$ at time $t_k$ with probability
$\frac{W(\C_k\to\C_{k+1})}{r(\C_k)}$ (See Figure~\ref{fig:timearrow}).
Note that contrary to the discrete time case, $\C_k$ and $\C_{k+1}$
are necessarily different. The time elapsed between two consecutive
jumps is a random variable: if the system arrives at time $t_0$ in the
configuration $\C_0$, the next move occurs at a time $t_1$,
distributed according to a Poisson law:
\begin{equation}
  \label{eqn:Poisson}
  \rho (t_1|\C_0,t_0)=r(\C_0) \ee^{-(t_1-t_0) r(\C_0)}
\end{equation}

\subsection{Large deviation functions}
\label{sec:ldf}

Let us consider an observable $A$ extensive in time, which can be
written as a sum along a history $\{\C_k\}_{0\leq k \leq K}$ of
elementary contributions $\alpha_{\C_k\C_{k+1}}$:
\begin{eqnarray}
  \label{eqn:defalpha}
  A = \sum_{0\leq k\leq K-1} \alpha_{\C_k\C_{k+1}}
\end{eqnarray}
This form is quite generic and most of the commonly studied observables
fall in this class. For instance, if $A$ is the overall current of a one
dimensional lattice gas, $\alpha_{\C\C'}$ is the
contribution of a single particle jump (See Section
\ref{subsec:SEP}). Moreover, to compute the average of a static observable ${\cal O}$ along
a history, one simply takes $\alpha_{\C\C'}={\cal O}(\C)$ and recover
$\left\langle {\cal O}\right\rangle=\frac{A}{t}$.

In equilibrium statistical mechanics, the difficult task of computing
the entropy can be conveniently substituted by the determination of
the free energy. Here, instead of working at fixed value of $A$
(microcanonical ensemble) to compute $\pi(A/t)$, one rather introduces
a parameter $s$ which fixes the average of $A$ (canonical
ensemble).$\,s$~is intensive in time and plays a role analogous to the
inverse temperature in equilibrium thermodynamics. This leads us to
introduce the dynamical partition function
\begin{equation}
\label{eqn:genfunc}
Z(s,t)=\left\langle \ee^{-s t a}\right\rangle\sim \ee^{t \psi_A(s)} \qquad\text{as}\qquad t\to\infty
\end{equation}
where the average $\langle\ldots\rangle$ is taken over all histories
between $0$ and $t$ and $a=A/t$ is intensive in time. The large
deviation function $\psi_A(s)$ of $Z(s,t)$ is the Legendre transform of
$\pi(a)$
\begin{equation}
  \psi_A(s)=\max_{a} \left[ \pi(a)-sa \right].
\end{equation}
Under quite general conditions \cite{ellis85}, this relation can be
inverted and one can get $\pi(a)$ knowing $\psi_A(s)$ through:
\begin{equation}
  \pi(a) = \max_{s} \left[\psi_A(s)+s a\right]
\end{equation}
To compute $Z(s,t)$, let us first write the
master equation obeyed by the joint probability $P(\C,A,t)$ of being
in configuration $\C$ with a value $A$ for the observable at time $t$:
\begin{equation}
  \partial_t P(\C,A,t) = \sum_{\C'\neq\C} 
  W(\C'\rightarrow\C) P(\C',A-\alpha_{\C'\C},t)\ - r(\C) P(\C,A,t)
\end{equation}
The Laplace transform $\hat P(\C,s,t) = \sum_A \ee^{-s A} P(\C,A,t)$
then evolves with
\begin{equation}  \label{eqn:evol_hatP}
  \partial_t \hat P(\C,s,t) = \sum_{\C'\neq\C} 
  W_s(\C'\to\C) \hat P(\C',s,t)\ - r(\C) \hat P(\C,s,t)
\end{equation}
where  the $s$-modified rates $W_s$ are given by
\begin{equation}
 \label{eqn:newrates}
 W_s(\C'\to\C)  = \ee^{-s\,\alpha_{\C'\C}}  W(\C'\to\C)
\end{equation}
From equation \eref{eqn:evol_hatP}, one sees that $Z(s,t)=\sum_\C \hat
P(\C,s,t)$  behaves at large time as $\ee^{\psi_A(s)t}$ where
$\psi_A(s)$ is the largest eigenvalue of a (not probability conserving)
evolution operator, which justifies the asymptotic behaviour in
\eref{eqn:genfunc}. The determination of the large deviation functions
$\psi_A(s)$ then amounts to the computation of this eigenvalue, which we
address in the next section.

\subsection{A cloning algorithm}
\label{subsec:algo}

Let us consider the {\em $s$-modified} Markov dynamics defined by the
rates $W_s$, whose evolution operator reads
\begin{equation}
  \left(\mathbb W_s\right)_{\C\C'} = W_s(\C'\to\C) - r_s(\C) \delta_{\C\C'}
\end{equation}
where
\begin{equation}
  r_s(\C)=\sum_{\C'}W_s(\C\to\C').
\end{equation}
The evolution of $\hat P(\C,s,t)$ can be written as
\begin{equation}
  \label{eqn:evolPhat}
  \partial_t \hat P(\C,s,t)= \sum_{\C'}  \left(\mathbb W_s\right)_{\C\C'} \hat P(\C',s,t) +[r_s(\C)-r(\C)] \hat P(\C,s,t)
\end{equation}
The corresponding dynamics alternates changes of configuration
determined by the $s$-modified rates and exponential evolution of the
``non-conserved probability'' $\hat P(\C,s,t)$ with rate
$r_s(\C)-r(\C)$ (corresponding respectively to the first and second
term of the r.h.s of~\eref{eqn:evolPhat}). More details are given in Appendix A. The evolution
of $Z(s,t)=\sum_\C \hat P(\C,s,t)$ is consequently represented by a
population dynamics a la Diffusion Monte Carlo~\cite{anderson}. Let us
consider $\NN_0$ clones of the system evolving in parallel with
the $s$ modified dynamics. At any change of configuration of any clone
$c_\alpha$,
\begin{enumerate}
\item[(1)] $c_\alpha$ jumps from its configuration $\C$
  to another configuration $\C'$ with probability $W_s(\C\to\C')/r_s(\C)$.
\item[(2)] The time interval $\Delta t$ until the next jump of $c_\alpha$
 is chosen from the Poisson law \eref{eqn:Poisson} of parameter
  $r_s(\C')$.
\item[(3)] The clone $c_\alpha$ is either cloned or pruned with a rate $\Y(\C')=\ee^{\Delta t(r_s(\C')-r(\C'))}$
  \begin{itemize}
  \item[a)] One computes $y=\lfloor \Y(\C')+\eps\rfloor$ where $\eps$ is uniformly
    distributed on $[0,1]$.
  \item[b)] If $y=0$, the copy $c_\alpha$ is erased.
  \item[c)] If $y>1$, we make $y-1$ new copies of $c_\alpha$
  \end{itemize}
\end{enumerate}
Such a cloning procedure modifies the total number of clones by a
factor $X=\frac{\NN+y-1}{\NN}$, which represents the exponential evolution
of $\hat P(\C',s,t)$. Finally, $Z(s,t)$ is simply given by the increase of the
population:
\begin{equation}
Z(s,t)=\frac{\NN(t)}{\NN_0}
\end{equation}
However, such an algorithm may result in an exponential growth or a
complete decay of the clones and we consequently add a $4^{\text{th}}$ step to maintain their number constant:
\begin{enumerate}
\item[(4)] If $y=0$, one clone $c_\beta\neq c_\alpha$ is chosen at
  random and copied, while if $y>1$, $y-1$ clones are chosen uniformly
  among the $N+y-1$ clones and erased. To reconstruct $Z(s,t)$, we
  keep track of all $X$ factors
\end{enumerate}
The \LDF $\psi_A(s)$ is then recovered from the long time behaviour of the product of
the cloning factors:
\begin{equation}
  \label{eq:main_result} 
  \frac{1}{t}\ln X_1 \ldots
  X_{\cal K}=\frac{1}{t} \ln \left\langle \ee^{-sta} \right\rangle \sim
  \psi_A(s) \qquad \text{as} \qquad {t\to\infty}
\end{equation}
where ${\cal K}$ is the total number of configuration changes among all the histories between 0 and $t$.

Another interpretation of such a population dynamics has been discussed
by Grassberger in \cite{Grassberger02} for the discrete time case and
we shall present it shortly. A formal integration of
\eref{eqn:evolPhat} leads to
\begin{equation}
\label{eqn:aveg}
\left\langle \ee^{-st a}\right\rangle=\left\langle   \ee^{\int_0^t r_s(\C(t))-r(\C(t)) dt} \right\rangle_s
\end{equation}
where the average $\left\langle\ldots\right\rangle_s$ is taken over
all trajectories from $0$ to $t$ of the $s$-modified dynamics. The
change of rates $W \to W_s$ can be seen a sequential importance
sampling, which favours histories relevant for the computation of
$Z(s,t)$. The exponential within the r.h.s. of \eref{eqn:aveg} can be
seen as a weight over trajectories. Along the simulation, clones with a high weight
$w_H$ are replaced by $w_H$ clones with a weight 1, while clones with
low weight $w_L$ are either deleted with probability $1-w_L$ or given
a weight $1$ with probability $w_L$. This is very close to the cloning
strategy followed at step $(3)$.

Let us finally note that one can define a new measure over the space of
trajectories, such that the average of an observable ${\cal B}$ reads
\begin{equation}
  \overline {{\cal B}^s}=\frac{\left\langle {\cal B} \ee^{-s t a}\right\rangle}{\left\langle \ee^{-s t a} \right\rangle}
\end{equation}
This is the measure observed in the simulations with a fixed number of
clones, as one can see that the denominator is nothing but the
unconstrained population at time $t$.

\subsection{ Thermodynamic integration}
\label{subsec:thermoI}
The direct computation of $\psi_A(s)$ is in general quite noisy, because of the
finiteness of the clones population. One can alternatively compute
$\psi'_A(s)$ and then integrate it. From the definition of $\psi_A\,$:
\begin{equation}
  \psi_A(s)=\frac{1}{t}\ln \left\langle \ee^{-s t a} \right\rangle 
\end{equation}
one gets by derivation
\begin{equation}
  \psi'_A(s)=-\frac{\left\langle a \ee^{-s t a}\right\rangle}{\left\langle \ee^{-s t a} \right\rangle}=-\overline{a^s}
\end{equation}
which is nothing but the average value of $a$ within the population of
clones. Finally,
\begin{equation}
\psi_A(s)=- \int_0^s \overline{a^r} dr
\end{equation}
Thanks to the integration, the noise is smoothed out.

\section{Three examples}
\label{sec:examples}

\subsection{The Symmetric Exclusion Process (SEP)}
\label{subsec:SEP}

We now apply the algorithm to the large deviations of the total
current $Q$ in the Symmetric Exclusion Process~\cite{spohn83} with
closed boundary conditions. The system is composed of $N$ particles
diffusing on a one-dimensional lattice of size $L$. Each particle can
jump with rate $1$ to any neighbouring site, provided it is empty.  The total current~$Q$ increases or decreases by $1$ at
every move, depending on the direction of the jump. Using the notation
\eref{eqn:defalpha}, $\alpha_{\C\C'}=1$ or $-1$ when a particle moves to its
right or to its left, respectively.

\begin{figure}[ht]
  \begin{minipage}{.5\columnwidth}
    \include{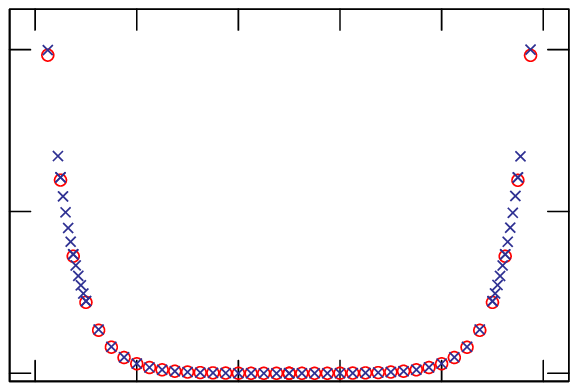}
  \end{minipage} 
  \begin{minipage}{.5\columnwidth}
    \include{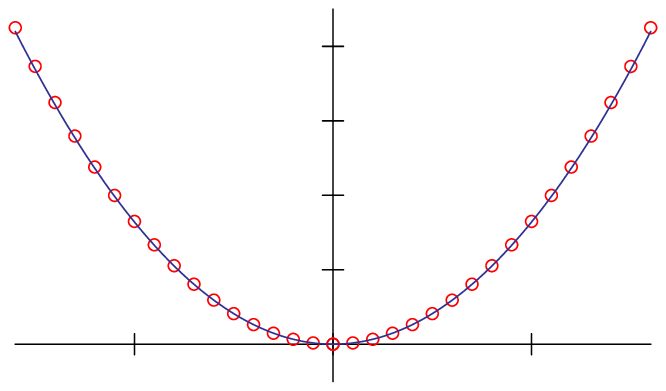}
  \end{minipage}
  
  \caption{Numerical evaluation of $\frac 1 L \psi_Q(s)$ for the Simple Exclusion Process
    ($N=200$, $L=400$). (a:~{\bf Left}) Comparison between direct 
    numerical measurement (blue crosses) and result from
    thermodynamic integration (red circles).  \\ (b:~{\bf Right})
    Comparison between numerical results (red circles) and analytical
    prediction~\eref{eq:dev_psiQ_SEP} valid for small $s$ (blue line).}
  \label{fig:SEP}
\end{figure}

As in other exclusion processes
~\cite{lebowitzspohn,derridalebowitzappert}, the \LDF $\psi_{Q}(s)$ is
extensive in the system size and we rather study the rescaled \LDF
$\frac{1}{L}\psi_{Q}(s)$ in the thermodynamic limit ($N$ and $L$ are
large, the density $\rho=N/L$ being fixed). Though on average the
total current is zero (the moves are symmetric), the variance is
finite and $\psi_{Q}(s)$ reads for small
$s$~\cite{bodineauderrida,spohn83}
\begin{equation} \label{eq:dev_psiQ_SEP}
  \frac 1 L \psi_{Q} (s) = \rho (1-\rho) s^{2} + \mathcal O(L s^{4})
\end{equation}
In this regime, the fluctuations are Gaussian and our algorithm yields
results in perfect agreement with the expansion~\eref{eq:dev_psiQ_SEP}
(see Figure~\ref{fig:SEP}b).  For larger values of $s$, we lack
an analytic expression for $\psi_{Q}(s)$ but the algorithm still
applies successfully. We found non-Gaussian fluctuations
(Figure~\ref{fig:SEP}a), which correspond to very large deviations of
the current. 

The direct measurement of $\psi_{Q} (s)$ is also compared with the
results obtained from  thermodynamic integration 
(Figure~\ref{fig:SEP}a). Both evaluations coincide within numerical
accuracy, though the latter requires duration more than one order of
magnitude smaller to converge.

\subsection{The Asymmetric Simple Exclusion Process (ASEP)}

\begin{figure}[t]
  \centering
  \begin{minipage}{.5\columnwidth}
    \include{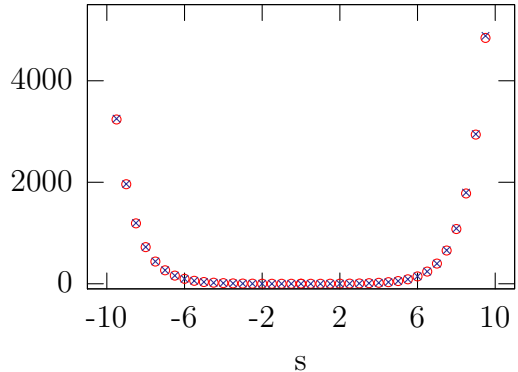}  
  \end{minipage}
  \caption{Plot of the \LDF $\frac 1 L \psi_{Q}(s)$ of the Asymmetric
    Simple Exclusion Process, for $L=400$ sites and $N=200$
    particles. The jump rates are $p=1.2$ and $q=0.8$, whence $E\simeq
    -0.2$. Blue crosses and red circles correspond to direct
    computation and thermodynamic integration respectively. The
    asymmetry appears when comparing the extreme points $s=\pm 9.5$.}
\label{fig:ASEPpsiQ}
\end{figure}
\begin{figure}[t]
  \begin{minipage}{.5\columnwidth}
    \include{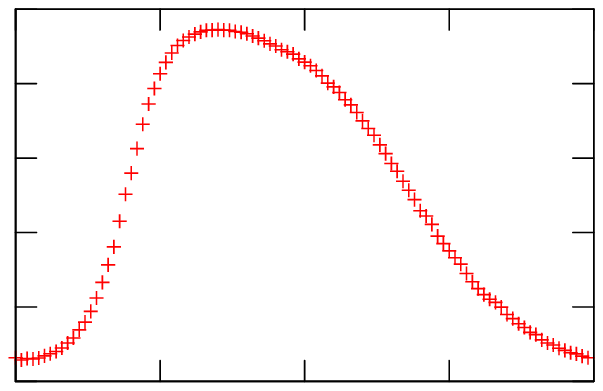}
  \end{minipage} 
  \begin{minipage}{.5\columnwidth}
    \include{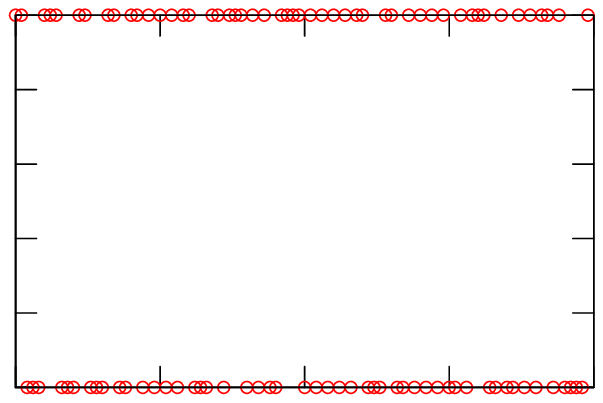}
  \end{minipage}
  \caption{(a:~{\bf Left}) Average profile $\rho$ for $s=0.3$. To
    minimise the overall current,
    the system develops an asymmetric shock, where only the front
    particles can jump easily.  \\
    (b:~{\bf Right}) A typical configuration for $|s|\gg E$. The
    particles are distributed almost uniformly.}
  \label{fig:ASEPconfigs}
\end{figure}

We now consider the large deviations of the total current $Q$ in a
non-equilibrium system, the Asymmetric Simple Exclusion
Process~\cite{spohn83} with closed boundary conditions. The system is
composed of $N$ particles diffusing on a one-dimensional lattice of
size $L$. Each particle can jump with rate $p$ to its left and $q$ to
its right, provided the arrival site is empty.  The total current~$Q$
is defined as in the previous section.  For $p\neq q$, a non-zero
steady current flows through the system.  We report in
Figure~\ref{fig:ASEPpsiQ} the large deviation function $\psi_{Q}(s)$.
One notes the symmetry around $E=\frac 12 \ln \frac qp\,$, guaranteed
by the fluctuation theorem.  The histories which contribute to
$\psi_{Q}(s)$ around $s\approx E$ ($Q \approx 0$) clearly display
shocks (Figure~\ref{fig:ASEPconfigs}{\bf a}) while a large current
($|s|\gg E$) is provided by uniform profiles
(Figure~\ref{fig:ASEPconfigs}{\bf b}).

\subsection{The Contact Process (CP)}
\label{subsec:CP}

We now turn our attention to the Contact Process~\cite{harris} in
one dimension.  The model is defined on a lattice of $L$ sites with
periodic boundary conditions.  Each site $i$ is either empty
($n_{i}=0$) or occupied by a particle ($n_{i}=1$). The dynamics is
defined as follows: particles annihilate with rate $1$, while empty
sites $i$ get occupied with rate
\begin{equation}
  W(n_i=0\to n_i=1)=\lambda (n_{i-1}+n_{i+1}) + h
\end{equation}
where $\lambda$ and $h$ are positive constants. Note the presence of a
spontaneous rate of creation~$h$. When $h=0$, the systems always
reaches an absorbing empty state in finite size
\cite{dickmanvidigal,deroulersmonasson}, while 
the steady state is active for $\lambda>1$  in the thermodynamic limit.

\begin{figure}
  \begin{minipage}{.5\columnwidth}
    \include{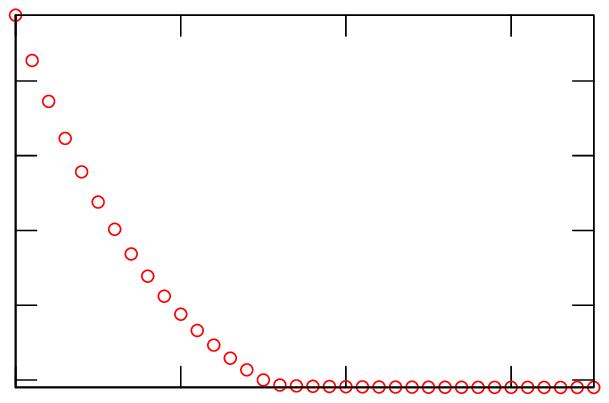}  
  \end{minipage}
  \begin{minipage}{.5\columnwidth}
    \include{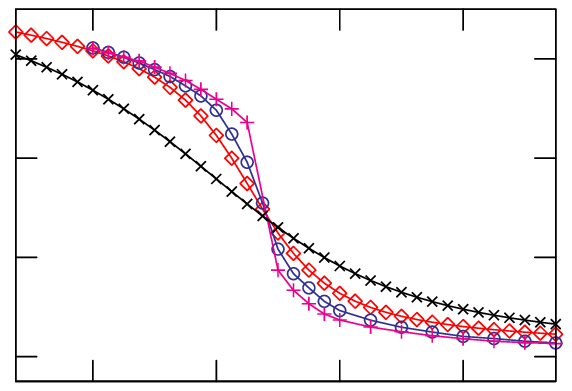}                
  \end{minipage}
  \caption{(a: {\bf Left}) Plot of the \LDF $\frac 1 L \psi_{K}(s)$
  associated to the number of events $K$ in the Contact Process in
  a field ($L=120$ sites) (b: {\bf Right}) The dynamical phase
  transition occurs at $s_{c}\sim 0.057$.  This is exemplified by 
  plotting $\psi_{K}'(s)=\frac 1 t\langle K\rangle_{s}$ for different 
  system sizes ($L=4$ in black, 8 in red, 15 in blue and 50 in magenta)} 
\label{fig:CP}   
\end{figure}

On the contrary, the addition of a small field $h$ leads to an active
equilibrium state for all $\lambda$. In the mean field version, the
equilibrium dynamics is still influenced by the presence of an
inactive state.  This can be seen through the study of the large
deviations of the number of events $K$, a quantity which simply counts
the number of configuration changes during an history of the system.
It was proved that $\frac 1L \psi_{K}(s)$ is non-analytic at a
critical value $s_{c}$, which goes to zero with $h$.  As pointed out in
section~\ref{sec:ldf} the \LDF $\psi_{K}(s)$ plays the role of a
dynamical free energy, whose non-analyticities are synonymous of
dynamical phase transitions.  In physical terms, this corresponds to the
existence of two distinct classes of
histories~\cite{thermo-formalism}: the active phase dominates the
steady state, while large deviations corresponding to $s>s_c$ are
dominated by inactive histories. For $s=s_{c}$, the two phases coexist
in a first-order fashion.  The question whether this transition is
still present in finite dimension is still open.

\medskip

Using our algorithm, we obtain evidence that this is indeed the case
in dimension $1$. In Figure~\ref{fig:CP}a, we plot the function
$\psi_{K}(s)$ for a large system size and for values of the parameters
$\lambda=3.5$, $h=0.1$. The two branches of the function correspond to
the two dynamical phases mentioned above.
As in the mean field version, the non-analycity appears as a jump in
the first derivative of $\frac 1 L \psi_{K}(s)$ in the thermodynamic
limit. Using the relation $\psi_{K}'(s)=-\frac 1 t \overline{K^{s}}$
(see section \ref{subsec:thermoI}), we can study the finite size scaling of  
$\frac 1 L \psi_{K}'(s)$ (Figure~\ref{fig:CP}).  The results
support the presence of a phase transition at $s_{c}\sim 0.057$.
Around $s_{c}$, the dynamics presents a superposition of ``more
active'' and ``less active'' histories, for which the continuous-time
approach is particularly helpful. 

\section{Conclusions}
\label{sec:concl}

We have presented a simple algorithm to evaluate large deviation
functions in continuous-time Markov chains without relying on any
time discretisation. We have shown on specific
examples that the method can be used successfully in systems where
the presence of different time scales renders the discrete-time
approach difficult.

In the context of quantum simulations, an approach to continuous-time
versions of Diffusion Monte Carlo was proposed by
Sylju{\aa}sen~\cite{syljuasen}, using a continuous formalism but a
discrete time implementation. One can expect that the algorithm
we have introduced could also be interesting in such context.

 
\ack This work has been supported in part by the French Ministry of
Education through the Agence Nationale de la Recherche's programme
JCJC/CHEF (VL). VL gratefully acknowledges the warm hospitality at the
Laboratoire de Physique et M\'ecanique des Milieux H\'et\'erog\`enes,
ESPCI, Paris, France, where part of this work was completed.  We thank
Christian Giardin\`a, Jorge Kurchan, Paolo Visco and Fr\'ed\'eric van
Wijland for useful discussions.

\appendix{}

\section{}

 An explicit formula for $Z(s,t)$ can be written as the sum
over $\C$ of the solution of \eref{eqn:evolPhat}:

\begin{eqnarray} 
  \label{eqn:sol_explicit} 
  Z(s,t) =& \sum_{K\geq 0} 
  \sum_{\C_1,\ldots,\C_{K-1},\C}
  \int_0^t d\mu^s_{t_K} \int_0^{t_K} d\mu^s_{t_{K-1}} 
  \ldots \int_0^{t_2} d\mu^s_{t_{1}} \ee^{-(t-t_K)r_s(\C)} \\
  &  \Y(\C_0)^{t_1-t_0} \frac{W_s(\C_0\rightarrow\C_1)}{r_s(\C_0)}\ldots
\Y(\C_{K-1})^{t_K-t_{K-1}}  \frac{W_s(\C_{K-1}\rightarrow\C)}{r_s(\C_{K-1})} \Y(\C)^{t-t_K}\nonumber
\end{eqnarray}
where
\begin{eqnarray}
  r_s(\C)  &= \sum_{\C'\neq\C} W_s(\C\rightarrow\C')
\end{eqnarray}
is the escape rate from the configuration $\C$ in the $s$-modified dynamics,
\begin{eqnarray}
d\mu^s_{t_k}&=dt_k\,r_s(\C_k) \exp[-(t_k-t_{k-1})r_s(\C_k)]
\end{eqnarray}
represents the probability of the time intervals between jumps, and
\begin{eqnarray}
\Y(\C_k)^{t_{k+1}-t_{k}}&=\ee^{(t_{k+1}-t_{k})(r_s(\C_k)-r(\C_k))}
\end{eqnarray}
is the exponential increase of $\hat P$ between $t_{k}$ and $t_{k+1}$.
We can read in this formula the direct expression of the cloning
factors used at the step $(3)$ of the algorithm.

\end{document}